\documentclass[journal=jacsat,manuscript=article]{achemso}

\usepackage[version=3]{mhchem} 
\usepackage{xcolor}
\usepackage[final]{pdfpages}

\usepackage{array}
\usepackage{multirow}
\newcolumntype{P}[1]{>{\centering\arraybackslash}p{#1}}
\author{Esteban~Rojas-Gatjens}
\affiliation{School of Chemistry and Biochemistry, Georgia Institute of Technology, Atlanta, GA, United~States}
\alsoaffiliation{School of Physics, Georgia Institute of Technology, Atlanta, GA, United~States}
\author{Hao~Li}
\affiliation{Department of Chemistry, University of Houston, Houston, Texas 77204, United~States}
\author{Alejandro~Vega-Flick}
\affiliation{School of Chemistry and Biochemistry, Georgia Institute of Technology, Atlanta, GA, United~States}
\author{Daniele~Cortecchia}
\affiliation{Center for Nano Science and Technology@PoliMi, Istituto Italiano di Tecnologia, Italy}
\altaffiliation{Current address: Dipartimento di Chimica Industriale Toso Montanari, Università di Bologna, 40136 Bologna, Italy.}
\author{Annamaria~Petrozza}
\affiliation{Center for Nano Science and Technology@PoliMi, Istituto Italiano di Tecnologia, Italy}
\author{Eric~R.~Bittner}
\affiliation{Department of Chemistry, University of Houston, Houston, Texas 77204, United~States}
\alsoaffiliation{Center for Nonlinear Studies, Los Alamos National Laboratory, United~States}
\email{ebittner@central.uh.edu}

\author{Ajay~Ram~Srimath~Kandada}
\affiliation{~Department of Physics and Center for Functional Materials, Wake Forest University, Winston-Salem, NC, United~States}
\email{srimatar@wfu.edu}

\author{Carlos~Silva-Acu\~na}
\affiliation{School of Chemistry and Biochemistry, Georgia Institute of Technology, Atlanta, GA, United~States}
\alsoaffiliation{School of Physics, Georgia Institute of Technology, Atlanta, GA, United~States}
\alsoaffiliation{School of Materials Science and Engineering, Georgia Institute of Technology, Atlanta, GA, United~States}
\email{carlos.silva@gatech.edu}

\title{Many-Exciton Quantum Dynamics in a Ruddlesden-Popper Tin Iodide}

\begin{document}

\begin{tocentry}
    \centering
    \includegraphics[width=0.5\linewidth]{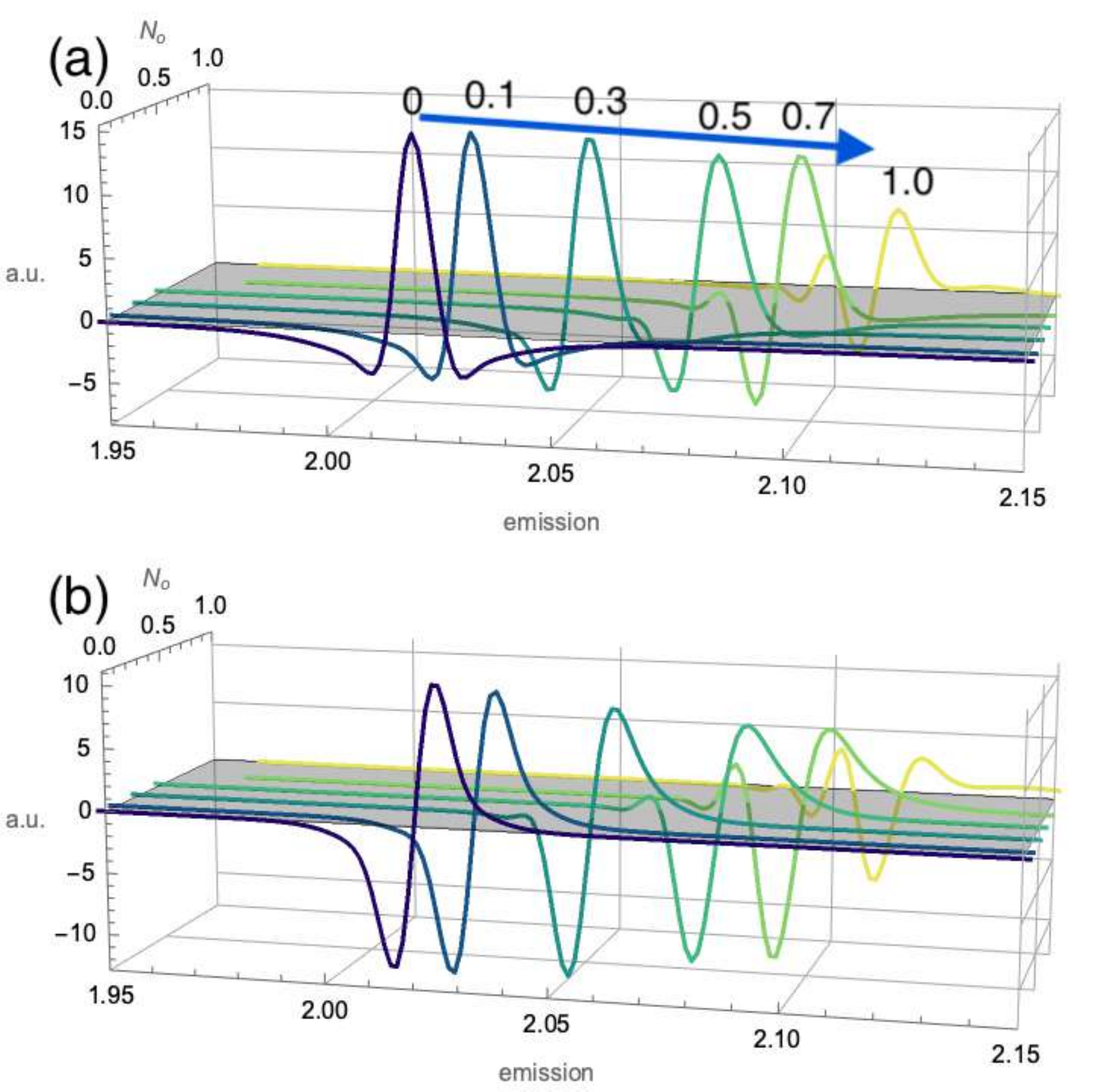}
\end{tocentry}

\newpage
\begin{abstract}
We present a study on the many-body exciton interactions in a Ruddlesden-Popper tin halide, namely \ce{(PEA)2SnI4} (PEA = phenylethylammonium), using coherent two-dimensional electronic spectroscopy.
The optical dephasing times of the third-order polarization observed in these systems are determined by exciton many-body interactions and lattice fluctuations. 
We investigate the excitation-induced dephasing (EID) and observe a significant reduction of the dephasing time with increasing excitation density as compared to its lead counterpart \ce{(PEA)2PbI4}, which we have previously reported in a separate publication [A.~R.~Srimath~Kandada~\textit{et~al.}, J.\ Chem.\ Phys.\ \textbf{153}, 164706 (2020)].
Surprisingly, we find that the EID interaction parameter is four orders of magnitude higher in \ce{(PEA)2SnI4} than that in \ce{(PEA)2PbI4}.
This increase in the EID rate may be due to exciton localization arising from a more 
statically disordered lattice in the tin derivative. 
This is supported by the observation of multiple closely spaced exciton states and the broadening of the linewidth with increasing population time (spectral diffusion), which suggests a static disordered structure relative to the highly dynamic lead-halide.
Additionally, we find that the exciton nonlinear coherent lineshape shows evidence of a biexcitonic state with low binding energy ($<10$\,meV) not observed in the lead system. We model the lineshapes based on a stochastic scattering theory that accounts for the interaction with a non-stationary population of dark background excitations.
Our study provides evidence of differences in the exciton quantum dynamics between tin- and lead-based RPMHs and links them to the exciton-exciton interaction strength and the static disorder aspect of the crystalline structure.
\end{abstract}

\section{Introduction}
Ruddlesden-Popper metal halides (RPMHs) are a unique class of semiconductors with a layered crystal structure, consisting of alternating metal-halide perovskite-like layers and long organic molecular moieties that separate them, forming a quasi-two-dimensional material. As a result of their structure, these materials host strongly bound excitons at ambient conditions~\cite{Blancon2018}, making them attractive for use in quantum optoelectronics applications such as photonic lasers~\cite{zhang20182d, zhang2018two, liang2019lasing, das2019perovskites, qin2020stable, lei2020efficient, liu2021solution, wang2022low, Alvarado2022} and exciton-polariton coherent emitters~\cite{Fieramosca2019, Polimeno2020}. 
In these quantum technologies, the quantum dynamics of excitons are deterministic since both population and coherence relaxation times are profoundly influenced by many-body exciton Coulomb correlations. 
A highly relevant example of such a phenomenon is excitation-induced dephasing (EID). 
EID arises due to \emph{instantaneous} incoherent elastic scattering between multiple excitons, leading to faster dephasing dynamics of the mesoscopic polarization than the low-density pure-dephasing limit and can be the dominant dephasing pathway at sufficiently high exciton densities. 
This process thus strongly determines optical lineshapes and limits homogeneous linewidths~\cite{Tokmakoof2000, Siemens2010, Graham2011}, which in turn govern the material's optical properties. 
A fundamental aspect of exciton many-body interactions in RPMHs is their strongly polaronic character~\cite{neutzner2018exciton, thouin2019phonon, Kandada2020}.
We have suggested that the primary photoexcitations in RPMHs are \textit{exciton polarons}, which are quasi-particles in which the anharmonic phonons and the Coulomb-bound electron-hole pairs are both integral components of the electronic eigenstates of the system. 
We have rationalized the presence of distinct excitonic resonances in the optical spectra of RPMHs is due to a family of distinct exciton polarons with distinct lattice dressing.~\cite{thouin2019phonon}. 
Notably, some researchers interpret the spectral structure as a vibronic progression~\cite{Straus2016, Dyksik2021_2, Straus2022, Kahman2021, Urban2020}, as opposed to our interpretation of distinct exciton polarons~\cite{Kandada2020}. In our previous study on \ce{(PEA)2PbI4}, we observed that the EID signatures are different for each of the resonances. If they belong to the same electronic excited manifold, then the many-body scattering should not vary significantly. Instead, we find that the polaronic dressings of excitonics present significant yet distinct consequences on the many-body quantum dynamics in RPMHs as observed through EID. 

%
%
Signatures of EID evidently manifest in the nonlinear coherent optical lineshapes~\cite{Li2006, Bristow2009, Paul2022}. 
A non-interacting coherent population typically results in a characteristic \textit{absorptive} lineshape in the real part of the two-dimensional coherent spectrum, composed of a symmetric, positive feature along the antidiagonal. 
The imaginary component of the spectrum concurrently exhibits a \textit{dispersive} lineshape that has a positive slope along the anti-diagonal. 
More precisely, for a two-level system, the rephasing response can be written as~\cite{Tokmakoof2000, mukamel1995principles}
\begin{equation}
    R_2(\omega_1,\omega_3) \propto
    \frac{1}{i(\omega_1+\omega_0)-\gamma}
     \frac{1}{i(\omega_3-\omega_0)-\gamma}, 
     \label{Eq:Reph_1}
\end{equation}
where $\omega_0$ is the transition frequency and $\gamma$ is the lorenzian width due to homogeneous dephasing. 
The expected spectrum of the antidiagonal line of the two-dimensional spectrum, centered about $\omega_0$ and represented as $\omega_1=-\omega_0+\delta$ and $\omega_3 = \omega_0 + \delta$, can be written as
\begin{equation}
    R_2(-\omega_0+\delta,\,\omega_0+\delta) = \frac{\gamma^2-\delta^2}{(\gamma^2+\delta^2)^2} + 2i \frac{\gamma\delta}{(\gamma^2+\delta^2)^2}.
    \label{Eq:Reph_2}
\end{equation}
The above expression clearly shows that the real part of the rephasing signal \emph{must} be an even function under reflection across the anti-diagonal while the imaginary component \emph{must} be an odd function. 
Moreover, it indicates that the imaginary part \emph{must} have a positive slope at $\delta = 0$ and the real part is positive at $\delta = 0$.  
Notably, this is a mathematically exact result for the rephasing signal and should be expected for any isolated transition with the homogeneous broadening of $\gamma$.  

In the case of RPMHs, previous observations have shown that the coherent nonlinear response deviates from the expected lineshape~\cite{Kandada2020Stochastic}. Instead, the real part of the spectrum appears to be an odd function with a positive slope along the anti-diagonal line, while the imaginary component is clearly absorptive. Such a reversal of the expected lineshapes of the real and imaginary components is indicative of an additional \textit{phase-shift} in the nonlinear response function of the sample. This shift has been previously interpreted as a consequence of excitation-induced effects by Cundiff and coworkers, based on a phenomenological model~\cite{Li2006, Turner2012}. 
We generalized this model by explicitly considering stochastic scattering of optically accessible excitations with a background population of dark excitons at non-zero momenta that leads to the overall phase-shift in the nonlinear response function~\cite{Kandada2020Stochastic}.

Our work establishes that scattering of the coherent exciton population with background excitations, a major portion of which are not optically inaccessible, is responsible for the observed relatively short coherence lifetimes in \ce{(PEA)2PbI4}. 
These scattering processes manifest even at low excitation densities, but they reduce substantially within hundreds of femtoseconds, much faster than the overall loss of the background population, but within the expected timescale of polaronic interactions. 
We hypothesized that the excitation-induced dephasing dynamics are limited by the dynamic screening of the many-body Coulomb interactions by the ionic lattice~\cite{Kandada2020Stochastic}.
The critical role of the lattice and the exciton polaron hypothesis, although suggestive, are yet to be robustly substantiated. 
A convenient experimental parameter that may enable further exploration of this problem is through metal cation substitution without majorly disrupting the structural composition of the lattice. This can be achieved in the \ce{PEA} derivatives in which the lead and tin counterparts have been shown to have similarities from a structural point of view~\cite{Mitzi1996}.

Although the effect of metal cation substitution from Pb to Sn and Ge has been extensively studied from a structural point of view~\cite{Mitzi1996, Knutson2005}, the changes it may induce in the electronic dynamics remain unclear. Firstly, the choice of the metal ion is expected to modulate the strength of spin-orbit coupling~\cite{folpini2023}, leading to consequences in the splitting energies between multiple conduction bands, composed of p-orbitals of the metal cation. Changes in the Rashba-Dresselhaus splitting at the band edge have also been predicted~\cite{krach2023emergence}. Secondly, moving from the larger \ce{Pb} ion to the smaller \ce{Sn} (or \ce{Ge}) can be expected to change the dynamic nature of the lattice due to enhanced organic-inorganic interactions, lattice polarizability, and consequently, the excitonic properties~\cite{Hansen2022, Kahman2021, Dyksik2021, Fu2021}. Both the above-stated consequences of the metal cation substitution can lead to pertinent changes in the many-body interactions and consequently, the excitation-induced dephasing dynamics.

In this article, we present a study on the effect of metal cation substitution on the excitonic Coulomb interactions in hybrid RPMHs.
We performed two-dimensional coherent electronic spectroscopy on a prototypical tin-based RPMH, \ce{(PEA)2SnI4}, to characterize the excitonic response. 
Specifically, we observed two distinct exciton signatures with different complex lineshapes and different EID behavior. 
To quantify the EID interaction parameter, we studied the linewidth dependence on excitation density. 
We also developed a microscopic interpretation of the phase-scrambled lineshapes using a stochastic scattering model and a biexciton excitation pathway, which helped us understand the differences between the two exciton polaron states. We further compare our findings with our previous work on the lead counterpart, \ce{(PEA)2PbI4}, and suggest that the differences in the nonlinear response could be attributed to their distinct degrees of static disorder. 
Overall, our results shed light on the effect of metal cation substitution on the excitonic response of RPMHs and provide insights into the underlying physics of these interesting materials.

\begin{figure}
    \centering
    \includegraphics[width=0.4\linewidth]{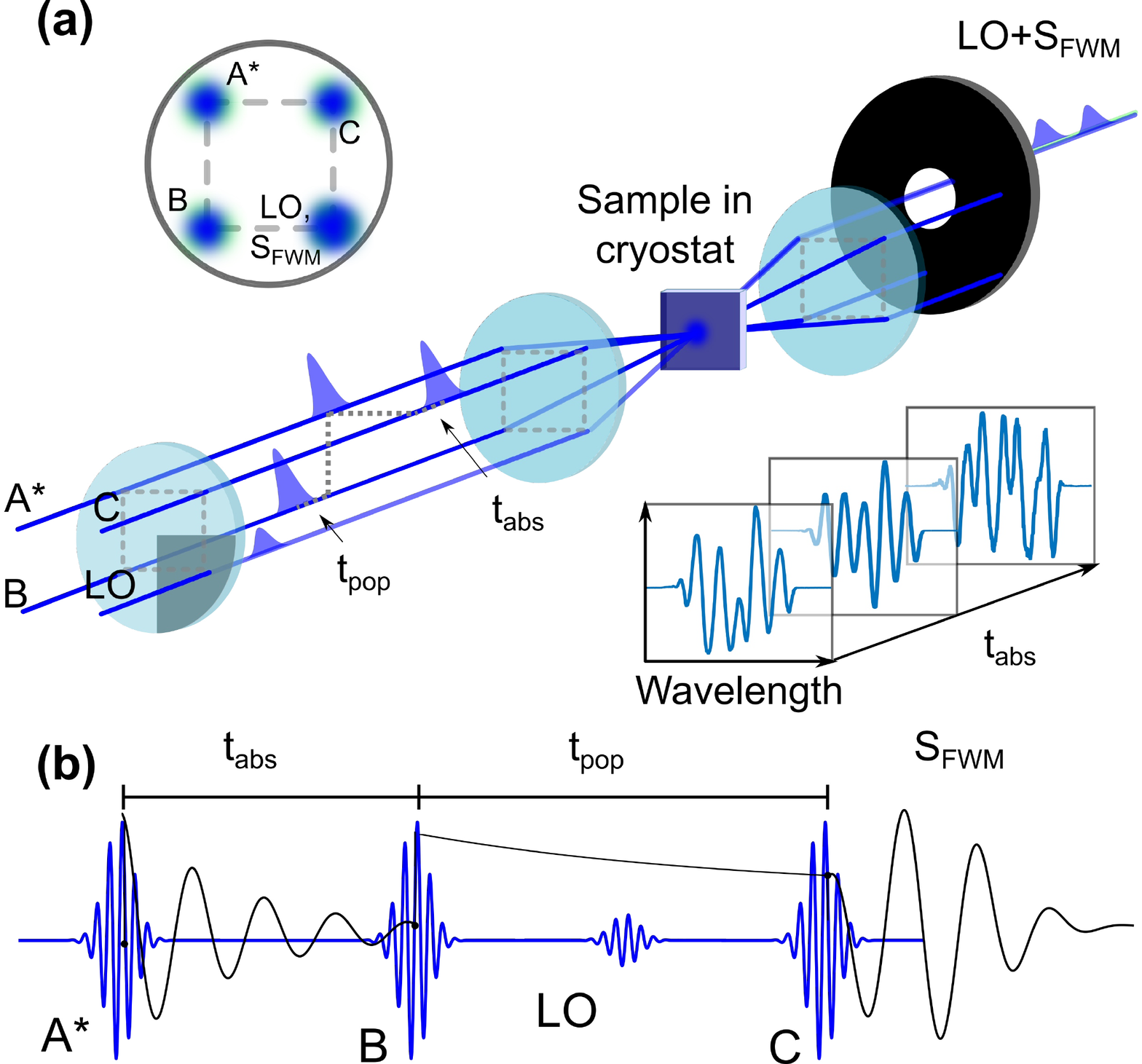}
    \caption{(a) Schematic representation of the two-dimensional coherent electronic spectroscopy in a ``Boxcar'' geometry. The experiment implements phase matching and time-ordering of the three light-matter interactions producing a third-order time-varying polarization that emits the coherent response, $S_{\mathrm{FWM}}$, along the fourth beam of the arrangement acting as a local oscillator (LO). $S_{\mathrm{FWM}}$ is detected by interference with a local oscillator. (b) Pulse train ordering corresponding to the rephasing experiment signal, with $k$-vector $\vec{k}_{FWM}=-\vec{k}_A+\vec{k}_B+\vec{k}_C$. Figure adapted with permission from ref.~\citenum{Thouin2019PRR}.}
    \label{fig:ExpScheme}
\end{figure}

\section{Results}

\subsection{Exciton Linewidth Broadening}

\begin{figure}
    \centering
    \includegraphics[width=\linewidth]{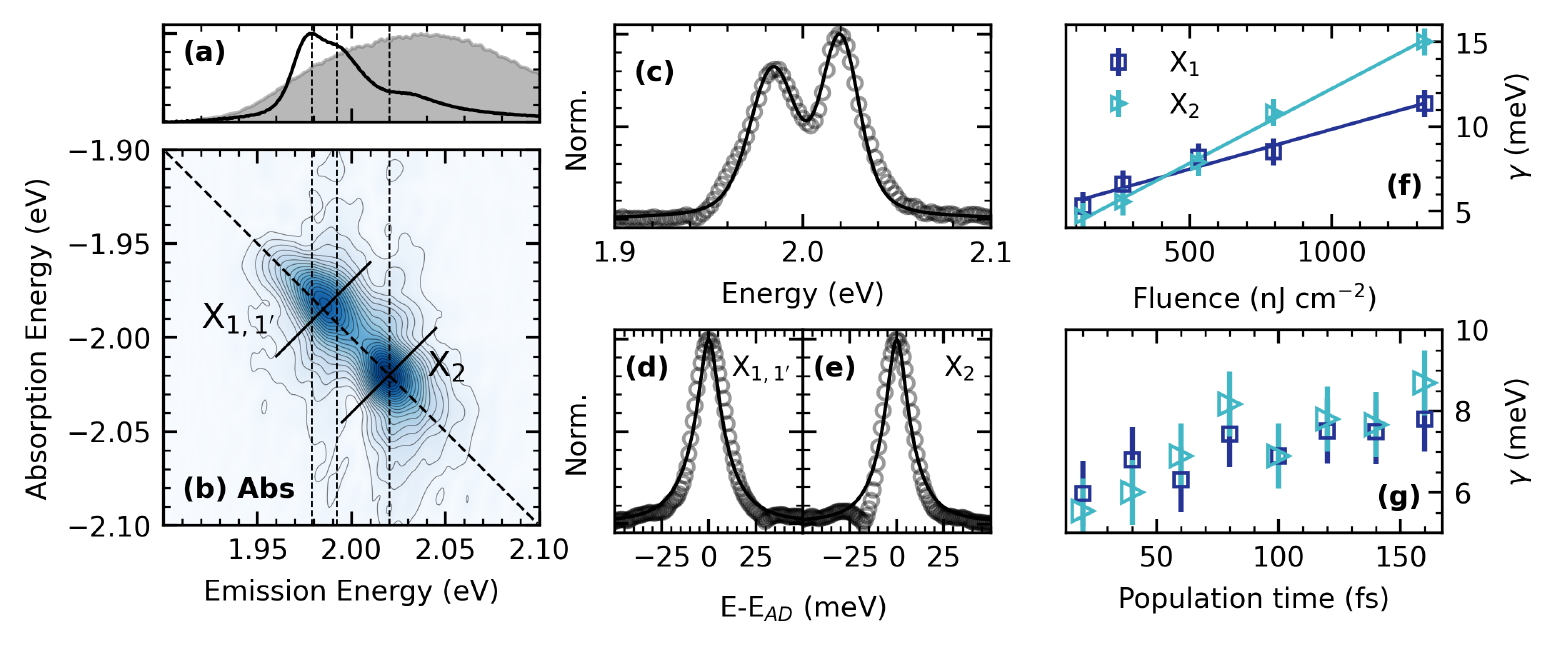}
    \caption{(a) Absorption spectrum of \ce{(PEA)2SnI4} (black line) measured at 15\,K and pulse spectrum (gray filled curve). (b) Absolute rephasing 2D coherent spectrum measured at 15\,K at a population waiting time of 20\,fs. The dashed lines mark the approximate position of the excitonic absorption at energies: 1.979, 1.992, and 2.020\,eV. (c) Normalized diagonal cuts and antidiagonal cuts at the diagonal energies of excitons X$_{1, 1^\prime}$ and X$_2$, (d) and (e) respectively, where $E-E_{AD}$ is the energy difference with the diagonal. (f) Fluence dependence and (g) population evolution of the homogeneous linewidth of $X_{1,1^\prime}$ and $X_2$}.
    \label{fig:2D_analysis}
\end{figure}

We first describe the main features observed in the linear and nonlinear absolute spectra. 
Three spectral features, labeled $X_1$, $X_{1^\prime}$, and $X_2$, can be clearly observed at the exciton energy in the linear absorption spectra, shown in Fig.~\ref{fig:2D_analysis}(a). 
In a two-dimensional Fourier transform experiment, a third-order polarization resonant with the exciton energy is generated by a sequence of three laser pulses incident on the sample in a ``boxcar'' geometry, shown in Fig.~\ref{fig:ExpScheme}. 
The coherent emission due to this induced third-order polarization is then detected through spectral interferometry with a fourth attenuated pulse (local oscillator, LO) co-propagated with the emitted field. 
Here, we focus on the analysis of the rephasing spectra where the emitted signal is acquired at $\vec{k}_{FWM}=-\vec{k}_A+\vec{k}_B+\vec{k}_C$.
In Fig.~\ref{fig:2D_analysis}(b), we show the absolute value of the rephasing 2D coherent spectrum measured with a fluence of 178~nJ\,cm$^{-2}$ per pulse. 
We observe two features along the diagonal corresponding to the energies of $X_1$ and $X_2$. Since $X_1$ and $X_{1^\prime}$ are not well resolved, we refer to the corresponding feature in the 2D spectrum as $X_{1,1^\prime}$. 
We also observe an off-diagonal cross peak between $X_{1,1^\prime}$ and $X_2$ with very low intensity. 
Note that the cross-peak has a $\pi$ phase-shift relative to the $X_{1,1'}$ feature observed in the real and imaginary components of the total correlation spectra, Fig.~S4. 
This suggests that the feature is due to an excited-state absorption pathway from an exciton state to a higher energy state common to both excitons. 
A similar scenario was described for the case of \ce{(PEA)2PbI4}~\cite{Thouin2018} where the higher-state is assigned to a biexciton, which has been a subject of study for several years in the community~\cite{Kondo1998, Fang2020, Ema2006}. 
In the case of \ce{(PEA)2SnI4}, we speculate that the excited-state absorption pathway involves a mixed biexciton between $X_{1,1'}$ and $X_{2}$. A detailed characterization of its binding energy and dephasing dynamics will be addressed in future work.

The elastic scattering events of excitons, e.g.\ exciton-exciton and exciton-phonon scattering, are characterized by their effect on the homogenous linewidth, $\gamma = \Gamma/2+\gamma_0$ where $\Gamma$ is the inverse of the exciton lifetime and $\gamma_0$ the pure dephasing term. 
The inhomogeneous nature of the semiconducting samples due to variations in the potential landscape leads to further broadening of the exciton excitation spectrum. 
From linear spectral measurement, we cannot rigorously separate the homogeneous and inhomogeneous contributions, we must therefore resort to nonlinear spectroscopy, particularly in materials such as RPMHs, which are highly dynamic~\cite{srimath2022homogeneous}. 
From an analysis of the diagonal and antidiagonal linewidths of the rephasing 2D coherent spectra, we can extract the inhomogeneous and homogeneous broadening contributions~\cite{Moody2015, Bristow2011}. 
Fig.~\ref{fig:2D_analysis}(c) shows a diagonal cut of the absolute rephasing map, while panels (d) and (e) show the antidiagonal cut. 
Since the homogeneous and inhomogeneous contributions are comparable and convoluted, the distinct contributions are extracted by simultaneously fitting the diagonal and anti-diagonal to the expressions derived previously by Siemens and Bristow~\cite{Siemens2010, Bristow2011} rather than simple Gaussian and Lorentzian functions.  
We obtained a homogeneous linewidth of $5.3\pm0.8$ and $4.7\pm0.8$\,meV for $X_{1,1^\prime}$ and $X_2$ respectively. 
The linear fluence dependence of the linewidth shown in Fig.~\ref{fig:2D_analysis}(f), clearly shows that excitons $X_{1,1'}$ and $X_2$ are both subject to excitation-induced dephasing due to inter-exciton interactions. 
The strength of the interaction is quantified by the slope of the fluence dependence fitted to the equation~\eqref{eq1}~\cite{Boldt1985, Moody2015}, where $\Delta_{ex}$ is the interaction parameter and $n$ is the excitation density. 
\begin{equation}
    \gamma \left(n\right) = \gamma_0 + \Delta_{ex}\,n.
    \label{eq1}
\end{equation}
We obtain a $\Delta_{ex}$ of $2.4\times10^{-8}$ and $4.5\times10^{-8}$\,$\mu$eV~cm$^2$ for exciton $X_{1,1^\prime}$ and $X_{2}$ respectively.  
We highlight that these values are orders of magnitude larger than the interaction parameters obtained previously for \ce{(PEA)2PbI4}~\cite{Thouin2019PRR}.  
The observed difference in the exciton-exciton interaction parameter between \ce{(PEA)2PbI4} and \ce{(PEA)2SnI4} is striking. 
We have suggested previously that many-body interactions are possibly screened by the lattice dressing of the excitations. 
Such polaronic effects are not expected to change drastically, moving from lead to tin, enough to quench the screening over three orders in magnitude. 
We hypothesize that the increased interactions in the \ce{(PEA)2SnI4} are instead a consequence of the increased exciton dipole moment, quantified recently by Hanse~\textit{et~al.}~\cite{Hansen2022} by electroabsorption spectroscopy. 
They reported that the dipole moment in the tin system is seven times larger in comparison to the lead counterpart and attributed it to reduced dynamic disorder within the lattice. 
In the same context, efficient dielectric screening has been linked to a dynamic distortion due to stereochemical expression of $ns^2$ electron pairs~\cite{Fu2021, Wang2021}, when substituting \ce{Pb^{2+}} with \ce{Sn^{2+}} the distortion becomes less dynamic (more static) and might result in the stabilization of charge separated excitons~\cite{Wang2021}. The difference in excitation-induced dephasing between lead and tin might arise from differences in the dynamic nature of the structure.

Another distinction between the lead and tin systems can be observed in the evolution of the homogeneous linewidth ($\gamma$) with population waiting time (delay between pulses B and C). 
In the case of \ce{(PEA)2PbI4}, we observed a rather peculiar reduction in the linewidth with time, which we interpreted as non-stationary evolution of the interaction with the dark background population. 
Here, contrary to our earlier observation, we see that the linewidth increases, albeit moderately, with population waiting time for \ce{(PEA)2SnI4}, as shown in Fig.~\ref{fig:2D_analysis}(g). 
Such line broadening is typically attributed to spectral diffusion~\cite{Hegarty1985, Singh2016} with the photo-excited coherent population accessing the inhomogeneous energy distribution that stems from static lattice disorder. 
This is also quantified by the \textit{diagonal} linewidth of $\delta \omega = 10\pm 1$\,meV from the fits in Fig.~\ref{fig:2D_analysis}(c), which is much larger than the estimates for \ce{(PEA)2PbI4} ($\delta \omega = 6.6\pm 0.1$\,meV). 
A larger static disorder, which may vary significantly from sample to sample and might also be related to \ce{Sn^2+} oxidation to \ce{Sn^4+}~\cite{Stoumpos2013}, does not allow us to visualize the time evolution in the screening of inter-excitonic interactions, which cannot be discounted in \ce{(PEA)2SnI4}. 

\subsection{Temperature dependent broadening}
\begin{figure}
    \centering
    \includegraphics[width=0.8\linewidth]{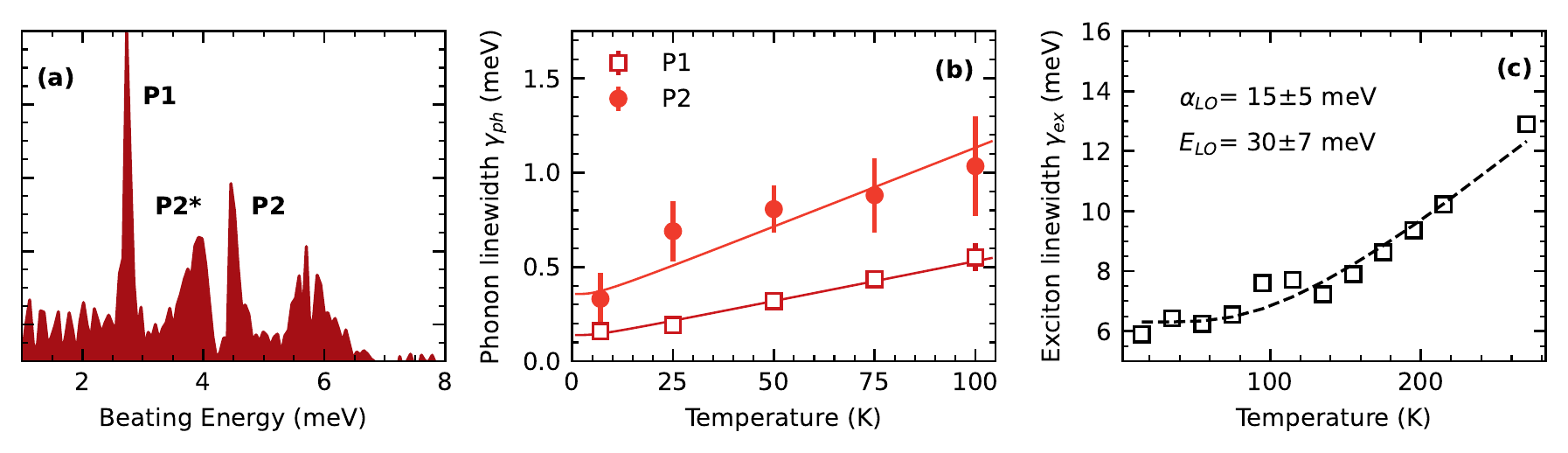}
    \caption{For \ce{(PEA)2SnI4}. (a) Cumulative RISRS spectrum measured at 7 K. (b) Temperature dependence of the dephasing time for the modes P$_1$ and P$_2$, the solid line is fit to the cubic overtone equation. (c) Temperature dependence of the X$_{1,1^\prime}$ homogeneous linewidth, obtained from the temperature-dependent linewidth analyzed by 2D coherent electronic spectroscopy.}
    \label{fig:raman}
\end{figure}

As  mentioned above, we observed an enhanced exciton-exciton interaction parameter for the case of \ce{(PEA)2SnI4}. 
We then are urged to study the anharmonicity and the exciton-phonon coupling, aiming to pinpoint the enhanced $\Delta_{ex}$ to differences in lattice-carrier coupling.
Resonance impulsive stimulated Raman scattering (RISRS) is an effective alternative to a cw resonance Raman, particularly for estimating the low energy phonon modes that dress photoexcitations in highly emissive materials with small Stokes shifts.
The evolution of a coherent superposition of phonon modes, generated through an impulsive excitation with an ultrashort optical pulse, modulates the complex refractive index. We observe such a  modulation as a time-oscillating differential transmission signal with characteristic lineshapes. An exhaustive description of the physical processes responsible for the generation of the coherent wavepacket and distinct experimental detection approaches can be found in Refs.~\citenum{Dhar1994RIRS, lanzani2007coherent} and for the case of RPMHs in Ref.~\citenum{Kandada2020}.
In previous work, we determined that the RISRS spectrum of \ce{(PEA)2PbI4} is primarily composed of two phonon modes with energies 2.61\,meV and 4.40\,meV~\cite{thouin2019phonon}. 
They were assigned to a pseudocubic axis and octahedral twist, and Pb–I–Pb bending, respectively.
In non-resonant impulsive stimulated Raman scattering, similar modes have been observed at 3\,meV and 6\,meV in \ce{(BA)2PbI4} and 3\,meV in \ce{MAPbI3}, the later assigned to anti-phase octahedra rotations~\cite{Guo2020}. 
We obtained the cumulative resonance Raman spectra of a \ce{(PEA)2SnI4} thin film from the modulation of the ground state bleach by Fourier transforming and binning the respective time traces,  Fig.~\ref{fig:raman}.a. We show the time traces and Fourier transform maps in the supplementary information.
We observe, unsurprisingly, a similar spectrum with an additional Raman mode below 4\,meV, labeled as P$_{2^*}$, as also reported recently in Ref.~\citenum{folpini2023}. 
We hypothesize P$_{2^*}$ arises due to the static distortion of \ce{(PEA)2SnI4} although it is hard to assign the origin of the static disorder due to the presence of self-doped domains the in the thin film due to progressive oxidation of the \ce{Sn^2+} to \ce{Sn^4+}. 

The temperature dependence of the phonon linewidth is directly related to anharmonic effects (phonon-phonon interaction) which becomes more significant as the phonon populations increase, shown in Fig~\ref{fig:raman}.b. 
We interpret the phonon-phonon interactions in terms of the overtone decay mechanism described by equation~\eqref{Cubic_overtone}~\cite{Vallee_GaAs, Klemens1966}. 
\begin{equation}
    \gamma_{ph} = \gamma_{0,ph}+\gamma_{anh}\,\left[1+\frac{2}{\exp\left(\hbar \omega/2k_BT\right)-1} \right].
    \label{Cubic_overtone}
\end{equation}
The $\gamma_{0, ph}$ corresponds to a temperature-independent rate which includes defect scattering~\cite{Maly2005} and $\gamma_{anh}$ is the anharmonic coefficient. We obtain for P$_1$ a $\gamma_{anh} = 0.016 \pm 0.1$\,ps$^{-1}$ and P$_2$ a $\gamma_{anh} = 0.05 \pm 0.01$\,ps$^{-1}$.
In order to compare across modes and materials we obtain the dimensionless parameter $\omega/2\pi\gamma_0$, shown in table~\ref{tbl:notes}. 
As observed for the case of \ce{(PEA)2PbI4}~\cite{Rojas2022} P$_1$ is less anharmonic than P$_2$ and in general we note that modes observed in tin are slightly more anharmonic than the lead counterpart. 
Independently, the temperature dependence of the exciton linewidth broadening provides the means of quantifying exciton-phonon coupling typically described by equation~\eqref{frohlich}.
\begin{equation}
    \gamma (T) = \gamma(0) + \alpha_{LO}\left(\frac{1}{\exp{E_{LO}/k_bT}-1}\right).
    \label{frohlich}
\end{equation}
In Fig.~\ref{fig:raman}.c, we show the dependence of the X$_{1,1^\prime}$ dephasing with temperature determined through coherent spectroscopy.
The intensity of X$_2$ decreases as temperature increases and could not be resolved unambiguously therefore the temperature dephasing is not shown in this work.
We estimate an interaction parameter, $\Delta_{ph}$, as $\Delta_{ph} = \alpha_{LO}k_B/E_{LO}$.
Its value is comparable to the one determined for \ce{(PEA)2PbI4}, summarized in table~\ref{tbl:notes}. 

\begin{table}
  \caption{Summary of parameters extracted in this work for \ce{(PEA)2SnI4} and previously obtained for \ce{(PEA)2PbI4} from Refs.~\citenum{Rojas2022, Thouin2019PRR}. We use the labels M$_1$ and M$_2$ from previous work to refer to the phonon modes in \ce{(PEA)2PbI4}. The parameters recover for the exciton-phonon coupling for \ce{(PEA)2PbI4} correspond to the exciton label as A in reference~\citenum{Thouin2019PRR}, the exciton with energy (2.37\,eV).}
  \label{tbl:notes}
  \begin{tabular}{P{1.5in}P{1in}P{1in}}
    \hline
    Parameters &  Value for \textbf{\ce{(PEA)2SnI4}} &  Value for \textbf{\ce{(PEA)2PbI4}} \\
    \hline
 $\omega_{ph}/2\pi\gamma_0$ (P$_1$/M$_1$) & 42$\pm$ 5 & -  \\
  $\omega_{ph}/2\pi\gamma_0$ (P$_2$/M$_2$) & 21 $\pm$ 9 & 35$\pm$6  \\
$\Delta_{ph}$ ($\mu$eV/K) & 44$\pm$20 & 70 $\pm$ 20 \\
$E_{LO}$ (meV) & 30$\pm$7 & 6.5$\pm$0.9\\
\hline
\end{tabular}
\end{table}

We note that equation~\eqref{frohlich} has been widely used for temperature-dependent linewidth analysis from linear photoluminescence measurements. 
However, it presents several limitations, specifically the model considers coupling with a single harmonic mode and does not consider short-range lattice coupling interactions. This is contrary to the strong anharmonicity reported in Ruddlesden-popper phases~\cite{Rojas2022, Menahem2021}
Additionally, in a previous perspective~\cite{Kandada2022}, we argued that the due to the inhomogeneous nature of Ruddlesden-Popper Metal-Halides the steady-state photoluminescence line shape is not strictly temperature independent if driven by diffusion-limited processes. 
The temperature-dependent population of dark states~\cite{Folpini2020, Dyksik2021, Kahman2021, Chandra2023} might also play a role in the temperature-dependent dephasing of the excitonic features.
For all these reasons, we restrain ourselves from attributing a robust physical meaning to the parameters extracted and from comparing with the many examples in the literature~\cite{Zhang2022, Hansen2022, Kahman2021}.

In this section, we characterized the exciton-phonon coupling and lattice anharmonicity. The results are not surprising, describing a very similar scenario for \ce{(PEA)2SnI4} compared to its lead counterpart, with just minor variations in the interaction parameters determined. 
This analysis discards a significant difference in the polaronic screening as the source of the very distinct exciton-exciton interaction parameter.
The enhanced exciton-exciton interaction parameter could be related to an extrinsic factor, for example, as described above due to enhanced strength of the exciton dipole moment induced by a disordered lattice~\cite{Hansen2022}, or exciton-carrier scattering due enhanced due to unintentional doping of tin perovskites~\cite{Treglia2022}.

\subsection{Complex Lineshape Analysis and Stochastic Scattering Modelling}

\begin{figure}
    \centering
    \includegraphics[width=0.85\linewidth]{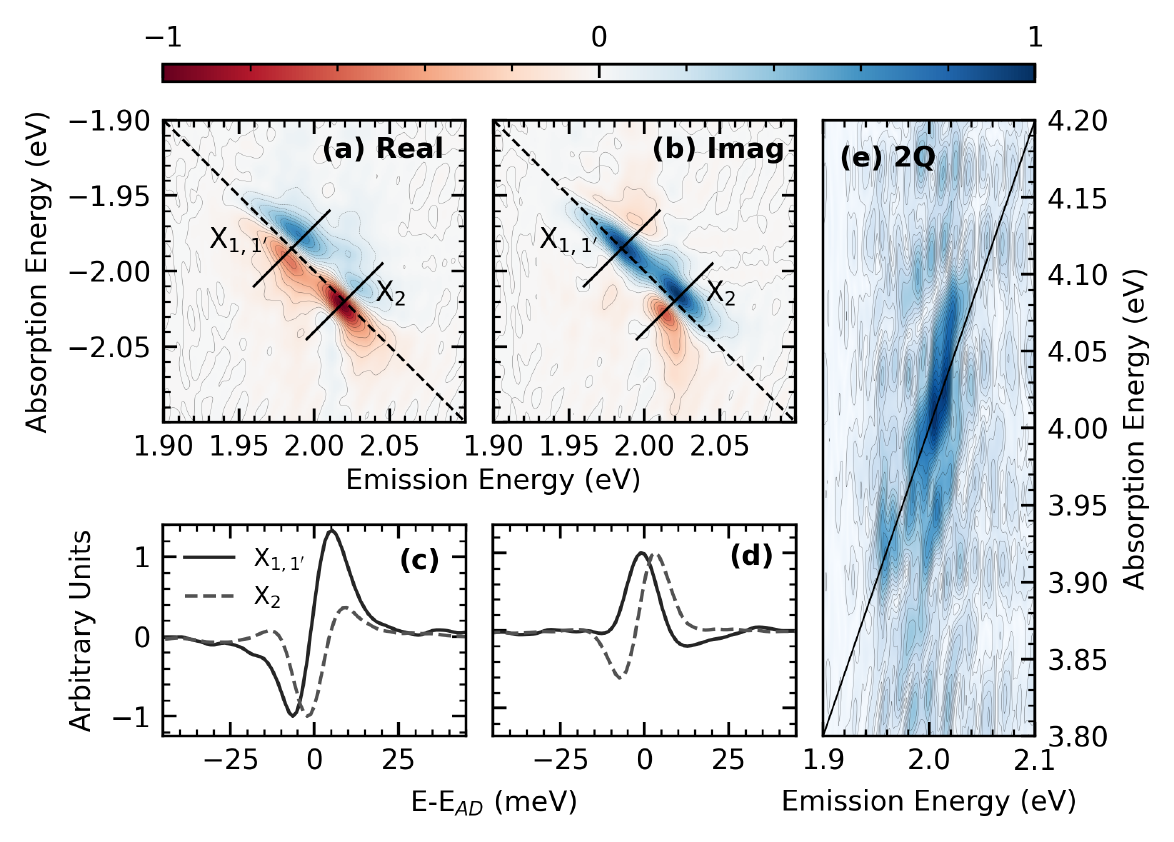}
    \caption{(a) Real and (b) imaginary components of the rephasing 2D coherent spectra measured at 15\,K at a population waiting time of 20\,fs, phased at 1.983\,eV. Antidiagonal cuts of the (c) real and (d) imaginary components showing the two main excitonic components $X_{1,1'}$ (solid line, black) and $X_2$ (dashed line, grey). (e) Two-quantum non-rephasing 2D coherent excitation absolute measurement of \ce{(PEA)2SnI4}.}
    \label{fig:2D_complex}
\end{figure}

The linewidth analysis of the absolute value of the rephasing spectrum, presented in the sections above, presents two important observations regarding the many-body interactions in \ce{(PEA)2SnI4}: (i) the interactions between excitons are much enhanced, possibly due to enhanced strength of the exciton dipole moment and (ii) the sample under study exhibits larger static lattice disorder. 
The first observation may be a direct consequence of the metal cation substitution, while the second one may also be a result of the variability in the sample fabrication conditions. While holding on to these observations, we now turn our attention to the details in the spectral line shape of real and imaginary components of the rephasing spectrum.
As noted in the introduction of this manuscript, the relative phase shifts in the nonlinear response, which manifest as deviations from expected absorptive and dispersive lineshapes in the real and imaginary components respectively, provide deeper insights into the origins of the many-body scattering~\cite{Bristow2008, Li2006, Kandada2020Stochastic}.  

The real and imaginary components of the rephasing 2D spectra are shown in Fig.~\ref{fig:2D_complex}(a) and (b) respectively.
We observe that the real part of the spectrum clearly dispersive at the energy of exciton $X_{1,1^\prime}$, with a positive slope along the anti-diagonal line and zero signal along the diagonal. The associated imaginary component is absorptive. 
The lineshape at $X{_{1,1^\prime}}$ is much similar to what we have previously reported for excitons in \ce{(PEA)2PbI4}, and much like in that case, we interpret it as a signature of excitation induced dephasing due to interaction with the background excitations. This is consistent with the fluence dependence of the homogeneous linewidth (see Fig.~\ref{fig:2D_analysis}(f)) that resulted in a larger exciton-exciton scattering parameter. 
The lineshape at $X_2$ in Fig.~\ref{fig:2D_complex}, however, does not follow this behavior. The real part at $X_2$ looks absorptive, but with a bit of asymmetry that shifts the positive peak to lower energies along the anti-diagonal line. The imaginary component looks dispersive, as expected for an isolated non-interacting system. This is absolutely not consistent with the estimated interaction parameter from the linewidth analysis for $X_2$. 
We consider that this discrepancy is not due to the failure of our photophysical model, but due to overlapping contribution from the excitation pathway to a biexcitonic state, which offsets the phase shift in the signal induced by many-body interactions.
This will be further discussed in the next section based on simulations of the nonlinear response using a stochastic scattering theory. 

Before dwelling into the simulations, we first independently verify the presence of a biexcitonic state associated with $X_2$ using a coherent two-quantum measurement (2Q)~\cite{Stone2009, Elkins2017}. The emitted field is collected at  $\vec{k}_{\mathrm{FWM}}=\vec{k}_B+\vec{k}_C-\vec{k}_A$ where the conjugated pulse interacts last. The outcome is a 2D map correlating the energy of the one quantum and two quantum excitations. The presence of the biexciton state can be confirmed by the appearance of a feature below the diagonal line in the 2D map, shown in Fig.~\ref{fig:2D_complex}(e). We clearly observe such a feature below 10\,meV at the energy of $X_2$ indicating a biexciton with very low binding energy. Notably, no such feature is observed at the energy of $X_{1,1^\prime}$. The binding energy is much lower than in \ce{(PEA)PbI4}~\cite{Thouin2018}, which will result in a substantial overlap of excitation pathways of exciton and its biexciton. 

We modeled the nonlinear response using the stochastic exciton scattering model developed in Refs.~\citenum{Kandada2020Stochastic}, \citenum{Li2020}, and \citenum{Li2022}. 
The model accounts for the non-stationary evolution of the population of dark background excitations, characterized by the initial average population density $N_0$ and variance $\sigma_N^2$. The non-stationary population is generated by each broadband excitation pulse and coupled to the optical mode via Coulombic interactions and evolves according to a stochastic differential equation corresponding to an Ornstein-Uhlenbeck process with damping rate $\Gamma$ and variance $\sigma^2$. 
The parameters in the model are related to the exciton-exciton interactions and in principle, can be determined by \textit{ab initio} or density functional theory methods.
Here, we treat the exciton-exciton interaction $V_0$ as adjustable parameters to examine how changes in the many-body interaction strength manifest in the nonlinear optical response. 
The most important components of the model are summarized in the supplementary information. 
We consider \textit{independent} transitions insofar as our theoretical model is concerned. 

\begin{figure}
    \centering
    \includegraphics[width=0.5\linewidth]{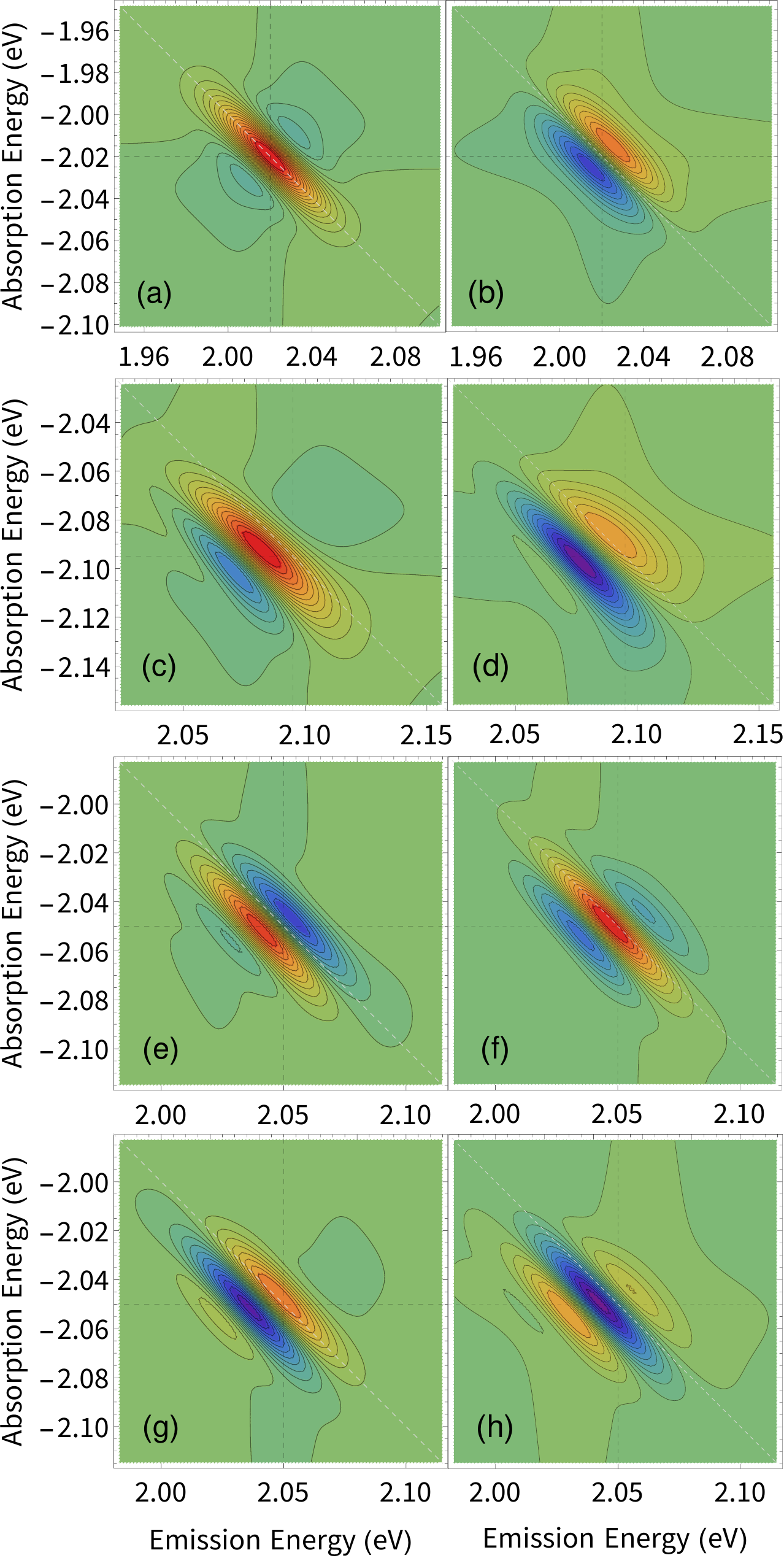}
    \caption{(a) Real and (b) imaginary components of a calculated rephasing spectrum with  background excitation density ($N_0 = 0$). 
    (c) Real and (d) imaginary components with a background excitation density ($N_0 = 0.2$), without the biexciton contribution.
    (e) Real and (f) imaginary components with background excitation density ($N_0 = 0.2$), with the biexciton contribution with $V_0 = $~5~mV.
    (g) Real and (h) imaginary components with background excitation density ($N_0 = 0.2$), with the biexciton contribution with $V_0 = -$5~mV.
    The background is modulated by an Ornstein-Uhlenbeck process of damping rate $\gamma=1.2~{\rm ps}^{-1}\approx 5~{\rm meV}$ and fluctuation variance $\sigma^2=25~{\rm ns}^{-1}$. }
    \label{fig:simulation}
\end{figure}

\begin{figure}
    \centering
    \includegraphics[width=0.5\linewidth]{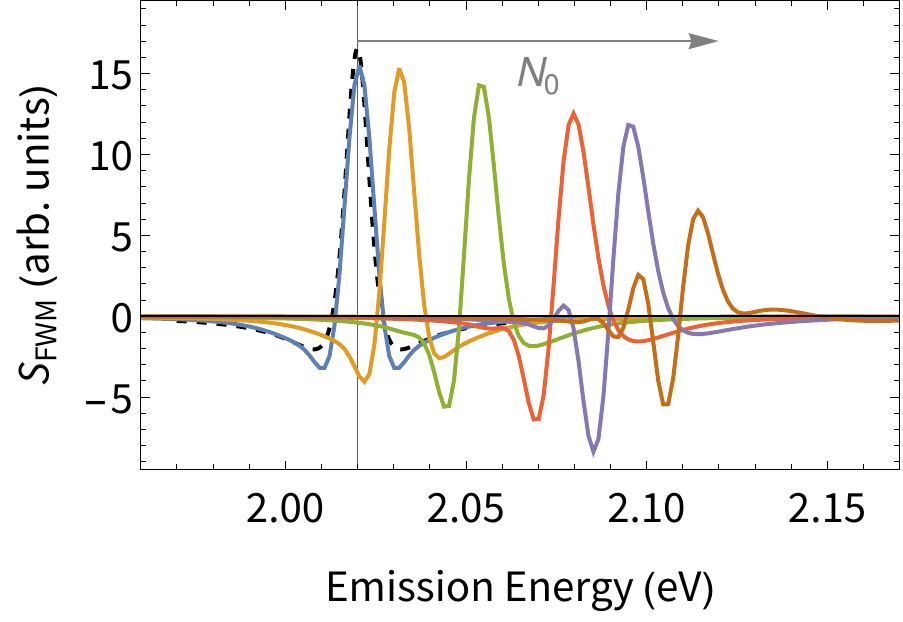}
    \includegraphics[width=0.5\linewidth]{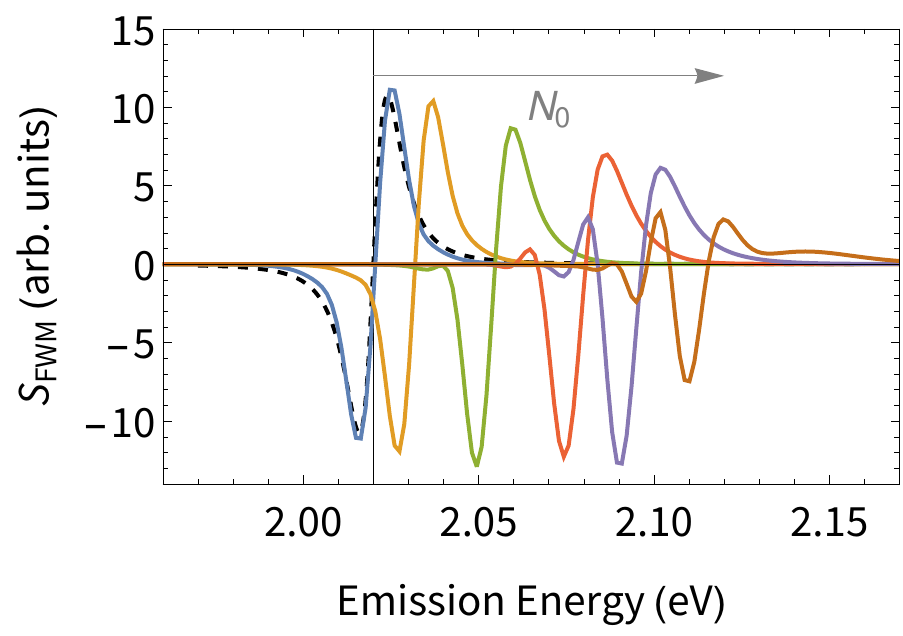}
    \caption{The effect of background exciton density $N_0$ on the homogeneous line shape of the coherent four-wave mixing signal $S_{\mathrm{FWM}}$. The anti-diagonal cut is taken at the strongest spectral peaks. $N_0=0$, $0.1$, $0.3$, $0.5$, $0.7$, $1.0$ from left to right. The black dashed curve is fitted from equation \eqref{Eq:Reph_2}, while the vertical grid line marks the energy of bare exciton. (Top) The real component of the 2D coherent spectra and (bottom) the imaginary component.}
    \label{fig:anti-diag-cut}
\end{figure}

When the initial population is set to zero ($N_0=0$), the model produces the rephasing spectra shown in Figs.~\ref{fig:simulation}(a) and (b), with symmetric lineshapes as expected for an isolated transition (c.f. Equation~\eqref{Eq:Reph_1}). 
Introducing a finite background population with $N_0 = 0.2$ produces the asymmetric lineshapes shown in  Fig.~\ref{fig:simulation}(c) and (d), reproducing qualitatively the experimental observation for  $X_{1,1^\prime}$. 
The asymmetry results from the additional phase factor that arises from the excitation-induced dephasing processes. 
This is evidenced in Figs.~\ref{fig:anti-diag-cut}(a) and (b), where we show the anti-diagonal cuts of the simulated rephasing spectra. We observe a clear evolution of the anti-diagonal of the simulated spectrum as we increase the interacting background population. Note that the real component transitions from an \textit{absorptive}-like lineshape when there is not an interacting background to a \textit{dispersive}-like lineshape as the $N_0$ increases. 
The imaginary component has the opposite behavior resulting in an apparent $\pi/2$ phase shift in the spectrum.
We note that the model can inherently account for the effect of exciton self-interaction (that leads to the biexciton) as a cross-peak whose position relative to the exciton diagonal feature is determined by the self-interaction amplitude $V_0$. 
This contribution is, however, not considered for the simulations shown in Figs.~\ref{fig:simulation}(c) and (d), given the lack of clear biexciton signature for the exciton $X_{1,1^\prime}$ in the two-quantum spectrum in Fig.~\ref{fig:2D_complex}(e).

To model the experimental lineshape corresponding to $X_{2}$, we incorporate the biexciton pathways by setting the exciton self-interaction to $V_0=\pm$5~meV. 
Note that a positive $V_0$ indicates a repulsive biexciton interaction and a negative $V_0$ indicates an attractive interaction.
Above, we mentioned that the biexciton pathway results in an emitted electric field with a phase factor of $\pi$ compared to the exciton pathway.
For a low biexciton binding energy, the diagonal exciton feature overlaps with the off-diagonal biexciton cross-peak,  producing a complicated lineshape. 
The overlap of two \textit{dispersive}-like lineshape with a phase difference of $\pi$ results in an apparent \textit{absortive}-like lineshape.
This is shown in Figs~\ref{fig:simulation}(e) and (f), with $V_0=$~5~meV, and Figs~\ref{fig:simulation}(g) and (h), with $V_0=-$5~meV.
Observe that including the biexcitonic term imparts an \textit{apparent} phase shift of $\pm \pi/2$ depending on the biexciton's binding energy. We refer to the phase shift as \textit{apparent}, since it is due to the overlap of two closely spaced spectral features, and thus the overall phase shift is not a real physical consequence. 
To summarize, the spectrum shown in Figs~\ref{fig:simulation}(c) and (g) qualitatively reproduced the experimental spectrum of $X_{1,1'}$ and $X_{2}$ respectively, shown in Fig.~\ref{fig:2D_complex}(a).
 

\section{Discussion}

As stated in the introduction, we conducted a study on the coherent nonlinear response of \ce{(PEA)2SnI4}, a tin-based RPMH, to comprehend the crucial role of the ionic lattice in many-body Coulomb interactions. Our investigation led to three significant differences in \ce{(PEA)2SnI4} compared to \ce{(PEA)2PbI4}: (i) a notably larger exciton-exciton interaction parameter $\Delta_{ex}$~\cite{Thouin2019PRR}, (ii) higher static disorder~\cite{neutzner2018exciton}, and (iii) lower biexciton binding~\cite{Thouin2018} energy, albeit with a higher transition cross-section.
Moreover, the anharmonicity of the lattice and exciton-phonon coupling of \ce{(PEA)2SnI4} are both comparable with those determined for \ce{(PEA)2PbI4} discarding a lack of polaronic screening as the origin of the observed differences.

The enhanced exciton-exciton interactions are discernible in the fluence-dependent exciton dephasing rates and the characteristic spectral lineshapes of the real and imaginary components of the rephasing spectra. We hypothesize that it can be attributed to the increased exciton dipole moment.  Interestingly, some researchers have suggested the possibility of a disorder-induced transition dipole moment enhancement~\cite{Horvath1995}, based on a comparison of the transition dipole moments of tin and lead-based RPMH~\cite{Hansen2022}. We observe a more static-disordered structure for the case of \ce{(PEA)2SnI4}, which can be observed in the population time evolution of the nonlinear spectra and the broad additional phonon mode in the Raman spectra. 
The description of the static disorder is significant as it impacts the excitonic properties of the material by inducing localized excitons through backscattering of the wavepacket from defect sites~\cite{Hegarty1985}.

Exciton-carrier scattering~\cite{Ashkinadze1995, livescu1988, Wu2015} is an additional scattering pathway that has not been addressed in this work explicitly. 
It possesses fundamentally a higher interaction strength than exciton-exciton interactions as it can be understood as a monopole interacting with a dipole instead of dipole-dipole interaction. For example, early work in GaAs single quantum wells determined exciton-carrier scattering is 8 times stronger than exciton-exciton interactions~\cite{Honold1989}. In the experiments presented here, we do not explore this interaction as we do not pump the free carriers with our pulse spectrum. Any carrier population from unintentional doping is not expected to result in the observed fluence dependence. Accordingly, we discount the contribution of exciton-carrier scattering to EID.  An interesting avenue to study this interaction may be available as the materials community further explores the doping~\cite{Treglia2022, Liu2022} of Ruddlesden-Popper metal halides.

We anticipate that the increased exciton-exciton interactions would also lead to an increase in the binding energy of the biexcitons, as the strength of the Coulomb coupling fundamentally governs both processes. However, contrary to such expectation, we observe that the biexciton binding energy in \ce{(PEA)2SnI4} is below 10\,meV, while it is about 50\,meV in \ce{(PEA)2PbI4}. This highlights the mechanistic differences between elastic many-body scattering of excitons and biexciton binding. The former can be interpreted as dipole-dipole scattering, with the interaction strength proportional to the \textit{physical} dipole's strength. In contrast, the biexciton results from four-particle (two electrons and two holes) correlations, and its binding energy is determined by the relative attraction and repulsion between charge carriers in the lattice. We emphasize that the biexciton feature can be observed in the rephasing spectrum of \ce{(PEA)2SnI4} at relatively low excitation density, while we required two orders of magnitude higher excitation density to observe it in \ce{(PEA)2PbI4}\cite{Thouin2018}. This can be attributed to the larger exciton-to-biexciton transition cross-section. However, it should be noted that in addition to lattice coupling, other physical phenomena, such as exchange interactions, may also influence the strengths of exciton-exciton interactions. Spin-selective lineshapes on excited state absorption features assigned to repulsive and attractive biexciton interactions have been observed in differential transmission spectroscopy following exciton spin dynamics of lead-bromide perovskite nanoplatelets\cite{Tao2020} and \ce{(PEA)2PbI4}\cite{Giovanni2018}. However, we have not explored this framework in our current work. Multidimensional spectroscopic experiments with circularly polarized light would significantly enhance our understanding of biexcitonic interactions~\cite{Stone2009}.
In our study, we demonstrated how phase shifts can occur as a result of exciton interactions with the background population. However, overlapping features, in this case, due to a biexcitonic excitation pathway, can also cause apparent phase shifts that are difficult to interpret without additional experiments, such as two-quantum excitation pathways.

\section{Perspective}

We employed two-dimensional Fourier transform spectroscopy to investigate the excitons in \ce{(PEA)2SnI4}. By performing a comprehensive analysis of linewidth and the complex lineshape, we identified differences in the many-body exciton interactions between the tin and lead samples. These differences provide insights into important chemical variables that could be manipulated to tune the exciton nonlinear interactions. We propose that subtle structural modifications in RPMH, where the lattice plays a deterministic role in the excitonic properties, could provide experimental access to distinct degrees of exciton-exciton interaction. This could be achieved by exploring the parameter space involving static and dynamic structure, dimensionality, chemical composition, and spin-orbit coupling strength. In this work, we observed that metal substitution resulted in an enhanced exciton interaction compared to its lead counterpart, and we observed an additional biexcitonic feature with low binding energy ($<10$\,meV).

\begin{acknowledgement}
The optical instrumentation was supported by the National Science Foundation (DMR-1904293). The experimental data collection, analysis, and the writing of corresponding manuscript sections by ERG, AVF, and CSA were supported by the National Science Foundation (DMR-2019444).
The work at the University of Houston was funded in part by the National Science Foundation (CHE-2102506, DMR-1903785) 
and the Robert A. Welch Foundation (E-1337).
The work by ERB was in part supported by the US Department of Energy through the Los Alamos National Laboratory. Los Alamos National Laboratory is operated by Triad National Security, LLC, for the National Nuclear Security Administration of U.S.\ Department of Energy (Contract No.\ 89233218CNA000001).
DC acknowledges support from by the European Union Horizon's 2020 research and innovation program under the Marie Skĺodowska-Curie grant agreement no. 839480 (PERICLeS).
\end{acknowledgement}

\begin{suppinfo}
Experimental procedures, additional characterization, and details regarding the model to simulate the nonlinear response can be found in the supplementary information.

\end{suppinfo}


\providecommand{\noopsort}[1]{}\providecommand{\singleletter}[1]{#1}%
\providecommand{\latin}[1]{#1}
\makeatletter
\providecommand{\doi}
  {\begingroup\let\do\@makeother\dospecials
  \catcode`\{=1 \catcode`\}=2 \doi@aux}
\providecommand{\doi@aux}[1]{\endgroup\texttt{#1}}
\makeatother
\providecommand*\mcitethebibliography{\thebibliography}
\csname @ifundefined\endcsname{endmcitethebibliography}
  {\let\endmcitethebibliography\endthebibliography}{}

\pagestyle{plain}


\section*{Supplementary material}

# 1 Experimental Methods

## 1.1 Sample preparation

The samples were prepared by mixing the precursors phenylethylammonium iodide (PEAI, Gratcell Solar Materials) and $SnI_2$ (TCI) in 2:1 stoichiometric ratio to form 0.05 M solutions in dimethylformamide (DMF, Anhydrous, Sigma Aldrich). The powders were left to dissolve overnight, then the solution was heated up to 100 °C for 1 hour before the deposition process. $(PEA)_2SnI_4$ polycrystalline thin films were spin-coated from the hot solution at 5000 rpm on fused silica substrates. The film was then annealed at 100 °C for 15 minutes on a hotplate. The entire process was performed under an inert atmosphere in a $N_2$ filled glove box. The films were encapsulated using a glass slide and a UV-cured resin. The resin was deposited only on the sample's edges to prevent direct contact with the central part of the sample where the measurements where performed. The perovskite film was masked during the UV-curing process to prevent its direct exposure to UV light.

## 1.2 Two-dimensional coherent electronic spectroscopy

In this work, we used our previously implemented scheme [1], [2], developed and described in detail by Turner and coworkers [3], to measure the 2D coherent spectra. The 1030-nm, 220-fs pulse train was generated in a commercial ultrafast laser system (Light Conversion Pharos) at a 100-kHz repetition rate. A portion of the laser beam was sent into a home-built third-harmonic-pumped non-collinear optical parametric amplifier. The pulses were individually compressed using a home-built implementation of a pulse shaper using a chirp scan [4]. The resulting pulse duration was 21 fs full-width at half-maximum, as measured by second-harmonic generation cross-frequency-resolved optical gating (SHG-XFROG). All measurements were carried out in a vibration-free closed-cycle cryostat (Montana Instruments).

## 1.3 Ultrafast Transient Absorption Spectroscopy

Using an ultrafast laser system (Pharos Model PH1-20-0200-02-10, Light Conversion) emitting 1030 nm pulses at 100 kHz, with an output power of 20 W and pulse duration of 220 fs. The measurements were carried out in a commercial set-up (Light Conversion Hera). The pump wavelength 3.06 eV was generated by feeding 10 W from the laser output to a commercial optical parametric amplifier (Orpheus, Light Conversion), while 2 W are focused onto a sapphire crystal to obtain a single-filament white-light continuum covering the spectral range 490–1050 nm for the probe beam. All measurements were carried in a vibration-free closed-cycle cryostat (Montana Instruments).

# 2 Ultrafast differential Transmission

## 2.1 Resonant Impulsive Raman Scattering

The electronic dynamics are subtracted from the differential transmission spectra using a polynomial fit. The resulting modulated response is then fast Fourier transformed at each probe-energy. The cumulative RIRS spectrum shown in the main text is obtained by binning the beating maps. The maps are shown in figure S2.

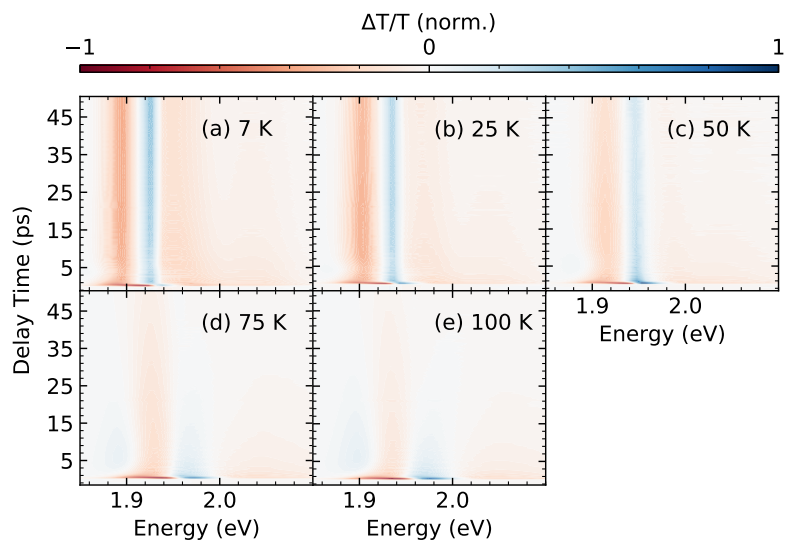

Figure S1: Time-resolved differential transmission spectra for $(PEA)_2SnI_4$ at temperatures (a) 7 K, (b) 25 K, (c) 50 , (d) 75 K and (e) 100 K.

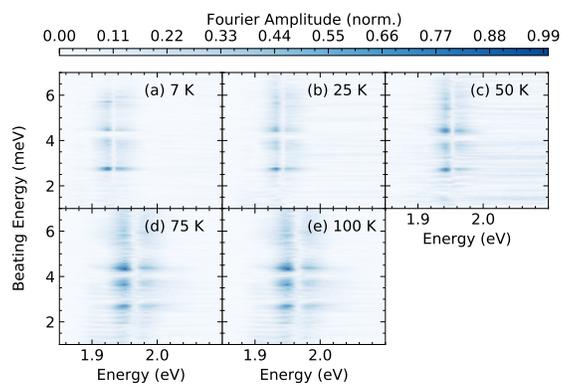

Figure S2: Resonant Impulsive Raman spectrum for $(PEA)_2SnI_4$ at temperatures (a) 7 K, (b) 25 K, (c) 50 , (d) 75 K and (e) 100 K.

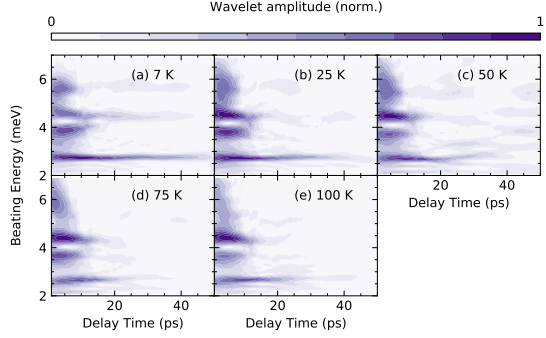

Figure S3: Wavelet transformation (CWT) spectrum of the binned time-domain RIRS data with a complex Morlet wavelet for $(PEA)_2SnI_4$ at several temperatures.

## 2.2 Wavelet Analysis Spectrograms

The CWT was implemented using the open-source wavelet transform software for Python under an MIT license. A complex Morlet wavelet was employed for the analysis, equation S1, with parameters B and C equal to 50 and 1 respectively. The wavelet transform is defined as equation S2, where $W(\tau, s)$ are the wavelet coefficients. The maps are shown in figure S3.

$$\psi(t) = \exp\left(-\frac{t^2}{B}\right) \exp\left(i2\pi Ct\right) \tag{S1}$$

$$W(\tau, s) = \frac{1}{|s|^{1/2}} \int_{-\infty}^{\infty} f(t)\, \psi^*\left(\frac{t-\tau}{s}\right)\, dt \tag{S2}$$

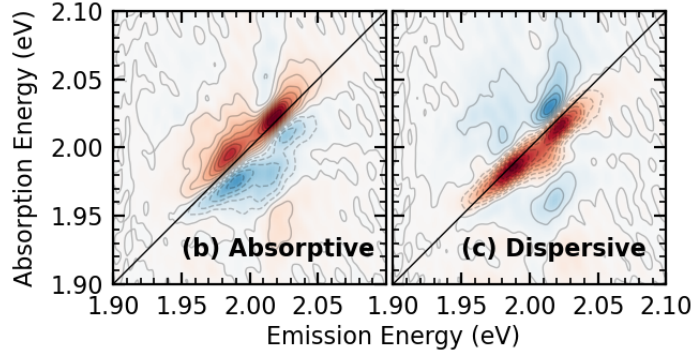

Figure S4: Total Correlation spectra (a) Absorptive and (b) dispersive components of the 2D spectra measured at 15 K at a population time of 20 fs, and phased at 1.983 eV.

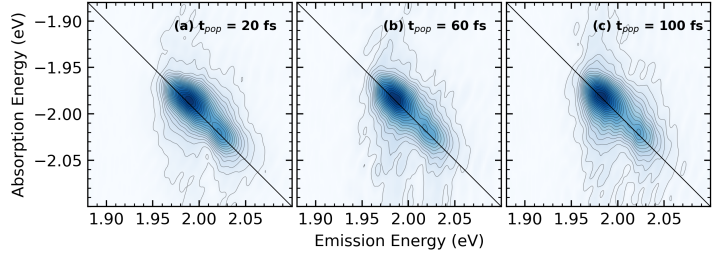

Figure S5: Absolute rephasing spectra at populations times (a) 20 fs, (b) 60 fs, and (c) 80 fs.

# 3 Multidimensional spectroscopy

## 3.1 Total correlation two-dimensional spectrum

The total correlation spectra is recovered as:

$$Abs = \Re(Rep) + \Re(Nonrep), \tag{S3}$$

$$Dis = \Im(Rep) + \Im(Nonrep). \tag{S4}$$

## 3.2 Linewidth analysis

Following the work by Siemens [5] and previous work [2]. We used the following expressions for extracting the homogeneous and inhomogeneous linewidths from the 1Q rephasing experiments.

$$S_{AD}(\omega_{ad}) = \left| \frac{\exp\left(\frac{(\gamma - i\omega_{ad})^2}{2\delta\omega}\right) \operatorname{erfc}\left(\frac{\gamma - i\omega_{ad}}{\sqrt{2}\delta\omega}\right)}{\delta\omega(\gamma - i\omega_{ad})} \right| \tag{S5}$$

Table 1: Summary of fit parameters obtained for two distinct spot in the $(PEA)_2SnI_4$

| Parameters | Position 1 | Position 2 |
|---|---|---|
| $\gamma_{E1}$ (meV) | 6.0 ±0.3 | 6.4±0.3 |
| $d\omega_{E1}$ (meV) | 9.9 ±0.4 | 10.2±0.3 |
| $\gamma_{E2}$ (meV) | 6.0 ±0.4 | 5.3±0.2 |
| $d\omega_{E2}$ (meV) | 6.8 ±0.4 | 11.2±0.3 |

$$S_D(\omega_d) = \sum_j \alpha_j \left| \frac{\exp\left(\frac{[\gamma - i(\omega_d - \omega_j)]^2}{2\delta\omega^2}\right)}{\gamma\delta\omega} \left[\text{erfc}\left(\frac{\gamma - i(\omega - \omega_j)}{\sqrt{2}\delta\omega}\right)\right] \right. \\ \left. + \exp\left(\frac{2\gamma i(\omega_d - \omega_j)}{\delta\omega^2}\right) \text{erfc}\left(\frac{\gamma + i(\omega_d - \omega_j)}{2\delta\omega}\right) \right| \quad (S6)$$

## 3.3 Fitting procedure

As noted from both the temperature dependence and fluence dependence the relative amplitude of $X_2$ makes the global fit unreliable as the inhomogeneous and homogeneous linewidth become correlated. We do global fit for measurements at two spots in the sample and obtain similar results, the inhomogeneous broadening parameter as well as the intensity changed in the new position. The global fit results are shown in table 1

## 3.4 Pulse duration characterization

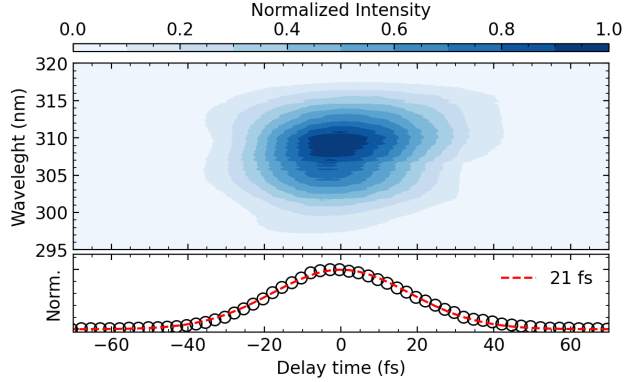

Figure S6: X-Frog characterization of the pulse duration fitted to a Gaussian function.

The beams are independently compressed using chirp-scan to a pulse duration of 21 fs FWHM, measured using cross-correlated second harmonic frequency-resolved optical gating (SH-XFROG) in a 10 μm thick BBO crystal placed at the sample position. A typical SH-XFROG trace is shown in Fig.S6.

## 4 Stochastic scattering model

A complete derivation of the stochastic theory is presented elsewhere [6]–[8]. Here, we summarized the most significant components of the work presented in this letter. The Hamiltonian in equation (S7) describe the scattering events modulated by $\gamma_1$, the coupling with a background population, $N(t)$, and $V_0$ is the exciton self-interaction energy.

$$H_0 = \omega_0 \hat{a}_0^\dagger \hat{a}_0 + \frac{V_0}{2} \hat{a}_0^\dagger \hat{a}_0^\dagger \hat{a}_0 \hat{a}_0 + \gamma_1 \hat{a}_0^\dagger \hat{a}_0 N(t). \tag{S7}$$

The time evolution of the operators is described as

$$\hat{a}_0^\dagger(t) = \exp\left(-i\left(\omega_0 + \frac{V_0}{2}\hat{n}_0\right)t - i\gamma_1 \int_0^t d\tau N(\tau)\right)\hat{a}_0. \tag{S8}$$

The two-dimensional spectra are simulated from response theory. For example, the expressions for a single excitation pathway are,

$$R_\alpha(\tau_1, \tau_2, \tau_3) = i^3 \mu^4 (n_0 + 1)^2 \exp\left[-i(\omega_0 + n_0 V_0)\sum_{j=1}^{3}(\pm)_j \tau_j\right] \\ \times \left\langle \exp\left[i\gamma_1 \sum_{j=1}^{3}(\pm)_j \int_0^{\tau_j} ds N(s)\right]\right\rangle, \tag{S9}$$

$$\left\langle \exp\left[i\gamma_1 \int_0^{\tau_j} N(s)ds\right]\right\rangle \approx e^{i\gamma_1 g_1(t) - \frac{\gamma_1^2}{2} g_2(t)}. \tag{S10}$$

The sign function $(\pm)_j$ takes "+" and "-" depending upon the sign of the photon wavevector entering or leaving the system. $N(t)$ is assumed to be a stochastic variable described by an Ornstein–Uhlenbeck process. The variance of the background population is given by $\sigma$ and the background relaxation rate is $\gamma$. The first and second cumulants are,

$$g_1(t) = \frac{N_0}{\gamma}\left(1 - e^{-\gamma t}\right), \tag{S11}$$

$$g_2(t) = \frac{\sigma^2}{2\gamma^3}\left[2\gamma t + 4e^{-\gamma t} - e^{-2\gamma t} - 3\right] + \frac{\sigma_{N_0}^2}{\gamma^2}\left(1 - e^{-\gamma t}\right)^2. \tag{S12}$$

[3] D. B. Turner, K. W. Stone, K. Gundogdu, and K. A. Nelson, "Invited article: The coherent optical laser beam recombination technique (colbert) spectrometer: Coherent multidimensional spectroscopy made easier," *Rev. Sci. Instrum.*, vol. 82, no. 8, p. 081 301, 2011. DOI: `10.1063/1.3624752`. [Online]. Available: `https://doi.org/10.1063/1.3624752`.

[4] V. Loriot, G. Gitzinger, and N. Forget, "Self-referenced characterization of femtosecond laser pulses by chirp scan," *Opt. Express*, vol. 21, no. 21, pp. 24 879–24 893, Oct. 2013. DOI: `10.1364/OE.21.024879`. [Online]. Available: `https://opg.optica.org/oe/abstract.cfm?URI=oe-21-21-24879`.

[5] M. E. Siemens, G. Moody, H. Li, A. D. Bristow, and S. T. Cundiff, "Resonance lineshapes in two-dimensional fourier transform spectroscopy," *Opt. Express*, vol. 18, no. 17, pp. 17 699–17 708, Sep. 2010. DOI: `10.1364/OE.18.017699`. [Online]. Available: `https://opg.optica.org/oe/abstract.cfm?URI=oe-18-17-17699`.

[6] A. R. Srimath Kandada and C. Silva, "Exciton polarons in two-dimensional hybrid metal-halide perovskites," *J, Phys. Chem. Lett.*, vol. 11, no. 9, pp. 3173–3184, 2020. DOI: `10.1021/acs.jpclett.9b02342`.

[7] H. Li, A. R. Srimath Kandada, C. Silva, and E. R. Bittner, "Stochastic scattering theory for excitation-induced dephasing: Comparison to the anderson–kubo lineshape," *J. Chem. Phys.*, vol. 153, no. 15, p. 154 115, 2020. DOI: `10.1063/5.0026467`. [Online]. Available: `https://doi.org/10.1063/5.0026467`.

[8] H. Li, S. A. Shah, E. R. Bittner, A. Piryatinski, and C. Silva-Acuña, "Stochastic exciton-scattering theory of optical line shapes: Renormalized many-body contributions," *J. Chem. Phys.*, vol. 157, no. 5, p. 054 103, 2022. DOI: `10.1063/5.0095575`. [Online]. Available: `https://doi.org/10.1063/5.0095575`.

# TOC Graphic

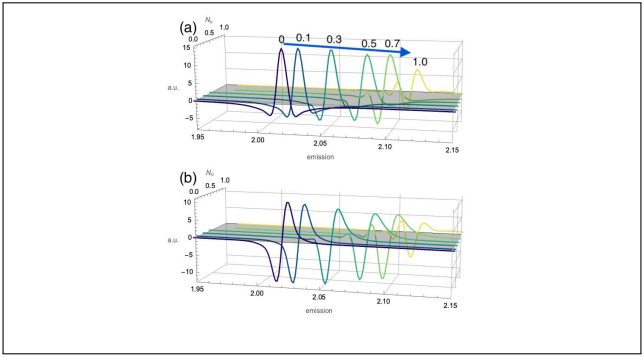



\end{document}